\documentclass[prl,twocolumn,superscriptaddress]{revtex4-1}

\usepackage{natbib}
\usepackage{graphicx}
\usepackage{color}
\usepackage{float}
\usepackage{mathrsfs}
\usepackage{amsmath,amssymb}

\begin{document}

\title{Gardner-like transition from variable to persistent force contacts in granular crystals}

\author{Lars Kool}
\affiliation{Laboratoire de Physique et M\'echanique des Milieux H\'et\`erog\`enes, ESPCI, Paris, France}
\affiliation{Department of Physics, North Carolina State University, Raleigh, North Carolina 27695, USA}

\author{Patrick Charbonneau}
\affiliation{Departments of Chemistry \& Physics, Duke University, Durham, North Carolina 27708, USA}

\author{Karen E. Daniels}
\affiliation{Department of Physics, North Carolina State University, Raleigh, North Carolina 27695, USA}

\date{\today}

\begin{abstract}
We report experimental evidence of a Gardner-like transition from variable to persistent force contacts in a two-dimensional, bidisperse granular crystal by analyzing the variability of both particle positions and force networks formed under uniaxial compression. 
Starting from densities just above the freezing transition, and for variable amounts of additional compression, we compare configurations to both their own initial state, and to an ensemble of equivalent, reinitialized states. This protocol shows that force contacts are largely undetermined when the density is below a Gardner-like transition, after which they gradually transition to being persistent, being fully so only above the jamming point. We associate the disorder that underlies this effect to the size of the microscopic asperities of the photoelastic disks used, by analogy to other mechanisms that have been previously predicted theoretically.
\end{abstract}


\maketitle

Granular materials differ from  elastic solids in their response to external forces: rather than homogeneously supporting an applied load, the forces are transmitted by a sparse percolating network of particles \cite{dantu1968etude, jaeger1996granular, radjai1998bimodal, howell1999stress, peters2005characterization}.
If interparticle contacts are allowed to break, and the granular material yields, the topology of the force network changes even if no particle-scale rearrangement takes place \cite{tighe2010force, kollmer2019betweenness}. By contrast, if contacts are preserved, cyclic (un)loading does not affect the structure of the force network.  While recent theoretical and numerical studies suggest the preservation of contacts might not coincide with the jamming transition~\cite{charbonneau2017glass, charbonneau2021memory}, it is yet to be experimentally verified whether such a distinction exists.

The distinction between the onset of contact memory and jamming is reminiscent of the critical transition reported for certain amorphous solids and crystals of slightly polydisperse particles \cite{gardner1985spin, kurchan2013exact, charbonneau2019glassy, berthier2019gardner, charbonneau2014fractal, tsekenis2021jamming, hammond2020experimental}. The associated Gardner transition is often depicted using an energy landscape roughened by a hierarchy of metastable basins. Outside of the Gardner regime, the energy scales are well-separated from the landscape roughness, and the system responds elastically \cite{biroli2016breakdown}. By contrast, within the Gardner regime, the landscape roughness gives rise to easier pathways to escape from marginally stable sub-basins and thus to minute  structural rearrangements (much smaller than the particle scale, ) 
that result in a different spatial distribution of contact forces at jamming \cite{jin2017exploring, charbonneau2019glassy}.

This landscape roughness in the Gardner phase also leaves a dynamical signature. Outside of the Gardner regime, the long-time mean squared displacement (MSD) $\Delta$ of the constituent particles plateaus at a value that depends on the particle cage size (and thus density/pressure for a hard sphere system) \cite{charbonneau2017glass}. By contrast, within the Gardner regime, particles can't effectively sample the landscape over accessible time scales, which results in a MSD that doesn't saturate with time.
Its asymptotically long-time value can nevertheless be estimated from the distance $\Delta_{AB}$ between two system copies, $A$ and $B$, that started from the same reference configuration at a density below the Gardner regime and then evolved along different stochastic trajectories. One can thus define the Gardner regime as the density for which   $\Delta < \Delta_{AB}$ at (sufficiently) long times. This was first shown experimentally in a granular glass former by  \cite{seguin2016experimental}, who captured a signature of Gardner physics in the dynamics of a vibrated, two-dimensional (2D), disordered packing of granular disks. More recently \cite{xiao2021probing} found signatures of Gardner physics in quasi-thermal (air-fluidized) star-shaped particles. However, the corresponding contact force network has not been observed experimentally, nor have the factors that control the distance of the Gardner transition to jamming been assessed \cite{charbonneau2019glassy}.

\begin{figure}
\includegraphics[width=\linewidth]{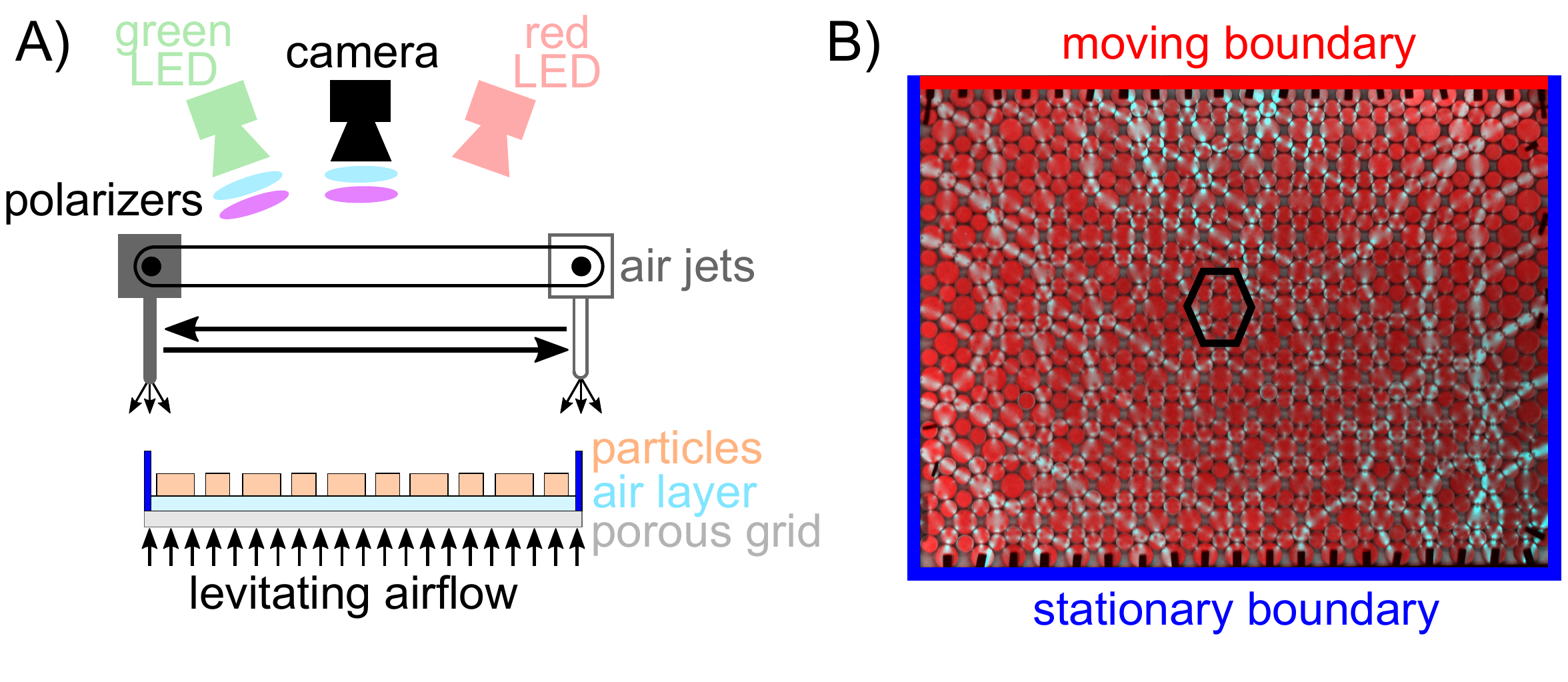}
\caption{A) Schematic of the experimental setup, side view, with the height of the air layer not to scale. B) Typical image (top view) from which the particle positions (red channel) and force transmission (cyan channel) are extracted. The hexagon marks an H1 unit cell \cite{likos1993complex}.}
\label{fig:Setup}
\end{figure}

In this Letter, we investigate the transition from variable to persistent contacts in a granular crystal (see Fig.~\ref{fig:Setup}). We find that this transition is strongly analogous to that predicted by Gardner physics, is clearly distinct from the jamming transition, and that the distance between the two appear here to be controlled by the scale of the microscopic asperities of the experimental disks.

\emph{Experimental setup--} Motivated by numerical studies of ultrastable glasses \cite{berthier2016growing} and polydisperse crystals \cite{charbonneau2019glassy} which successfully suppress particle-scale rearrangements to reveal the Gardner regime, we study marginally-stable states generated from a well-defined 2D crystalline packing. We selected the $H_1$ crystal symmetry, containing a unit cell of three large and six small disks (see Fig.~\ref{fig:Setup}), from among those identified by \cite{likos1993complex}, for having no basis vectors aligned with the compression axis. This choice thereby limits the putative contribution of low-energy, local buckling excitations~\cite{lerner2013low, charbonneau2015numerical, tsekenis2021jamming, charbonneau2015jamming}, and focuses the dynamics on the quenched disorder that arises through variability in particle size and  surface roughness~\cite{charbonneau2019glassy}. 
We found that this crystal successfully suppresses rearrangements; when one does occur, the system can readily be reinitialized. Although the resulting crystalline axes create an additional coordinate system that is neither orthogonal nor aligned with the natural axes of the apparatus (see Fig.~\ref{fig:Setup}), we are able to account for these effects during the data analysis. 

We performed our experiments on a single-layer packing of bidisperse photoelastic disks ($N_s = 507$ small disks with $d_s= 11.0$~mm and $N_l = 273$ large disks with $d_l = 15.4$~mm, Vishay PhotoStress PSM-4) with a reflective back layer levitated on a gentle layer of air forced through a porous grid; this setup has been previously described in~\cite{puckett2013equilibrating, bililign2019protocol}. By reducing basal friction, such that interparticle forces dominate, particles are free to explore their cages and sample available configurations under gentle perturbations. We randomize particle positions {\itshape within their cage} by sweeping a turbulent airflow across the upper surface of the packing (see Fig.~\ref{fig:Setup}a); time is measured in units of these $t_r = 20$~s randomization sweeps. We explore cage sizes and separations as a function of density $\phi$ by uniaxially compressing the system  in discrete increments of $\delta \phi / \phi = 6 \times 10^{-4}$, moving one boundary with a stepper motor. Each of the four boundaries are laser-cut from acrylic sheets, and the particles along the moving pair of boundaries are pinned in place to suppress large-scale crystal rearrangements. 

Particle positions and the network of interparticle forces were imaged using a single camera and two light sources: an unpolarized red LED light for the positions, and a circularly-polarized green LED light for the photoelastic visualization of stresses (see  Fig.~\ref{fig:Setup}b). 
We located the centroid of each particle using the convolution of the red channel of the image with a predefined mask matching the  particle size \cite{PeGS,daniels2017photoelastic}; this allows us to determine locations within $\approx 0.1$~px precision. Because we are studying well-defined crystal configurations, for which all particle displacements are at least an order of magnitude smaller than the particle size, we used a nearest-neighbor algorithm to both create particle trajectories and check that all particles had been detected. To minimize edge effects, particles within 2$d_l$ of the walls were discarded from the dataset, leaving $N_p=628$ particles for our analysis.

\emph{Results --}
We have adapted the protocol of \cite{seguin2016experimental} to identify transitions in the cage dynamics as a function of $\phi$, using overhead airjets to randomly promote cage exploration rather than supplying a global vibration of the bottom plate. We determine the cage dynamics at 20 different $\phi$, equally spaced between $\phi_\mathrm{min} = 0.8006 \pm 0.0002$ (limit of mechanical stability of the crystal) and $\phi_\mathrm{max} = 0.8162 \pm 0.0002$ (always slightly greater than $\phi_J$, guaranteeing that we have traversed the Gardner regime but without activating the out-of-plane buckling mode that develops deeper into the jammed phase). From the initial state ($\phi_\mathrm{min}$), the system is compressed to $\phi_\mathrm{max}$, and then decompressed step-wise and allowed to equilibrate for $t = 100t_{r} $ at each intermediate density. Upon reaching $\phi_\mathrm{min}$, the system is deemed \textit{reinitialized}. We performed a total of 10 initializations, shown schematically in~\cite{footSI}.

For each $\phi$, the cage separation distance $\Delta_{AB}$ is obtained by comparing particle positions between two different initializations, $A$ and $B$, taken at the same $\phi$:
\begin{equation}
	\Delta_{AB}(t;\phi) = \frac{1}{N_p} \sum_{i=1}^ {N_p}|\textbf{r}_i^B(t) - \textbf{r}_i^A(t)|^2
	\label{eq:Delta_AB}
\end{equation}
where $\textbf{r}_i^\alpha(t)$ is the position of particle $i$  at time $t$ in initialization $\alpha$. The cage size $\Delta$ (within a single initialization $A$) at a given $\phi$ is obtained from particle displacements after a long experimental time of $100 t_{r}$, according to
\begin{equation}
	\Delta(t;\phi) = \frac{D}{N_p} \sum_{i=1}^{N_p}|\textbf{r'}_{i,y}^A(t) - \textbf{r'}_{i,y}^A(0)|^2.
	\label{eq:Delta_AA}
\end{equation}
In both cases, the average at each $\phi$ over all runs is then calculated, denoted by $\langle\cdot\rangle$. Below the Gardner regime, $\langle\Delta\rangle$ and $\langle\Delta_{AB}\rangle$ should be equal at equilibrium.

For that to be the case in our system, two experimentally-motivated corrections are introduced.
First, even at $\phi_\mathrm{min}$ the MSD of a caged particle plateaus at longer times than are experimentally accessible; 
we measured this time to be $\approx 1000 t_r$, while our experiments can only reach $100 t_r$. Because we expect the relative ratio of these lengthscales to be constant at low $\phi$, we have rescaled our measurement of $\Delta$ by the ratio $D = 1.2$, our estimate of this ratio~\cite{footSI}.
Second, we observed that for $\phi > \phi_{G}$, histograms of the $(x,y)$ displacements displayed multiple distinct peaks, each aligned with the direction of one of the lattice vectors of the unit cell, rather than being azimuthally symmetric around zero, as observed for $\phi < \phi_G$. To correct for the biased motions introduced by the crystalline axes that give rise to these peaks, we applied a linear transformation to orthogonalize the system. Equation~\eqref{eq:Delta_AA} therefore defines $\Delta$ as the MSD of the Gaussian part of the displacement (along the $y'$-axis) in the orthogonalized system $r'_y$, and ignores the displacements along the more-complicated ($x'$) axis~\cite{footSI}.

\begin{figure*}
    \includegraphics[width=\textwidth]{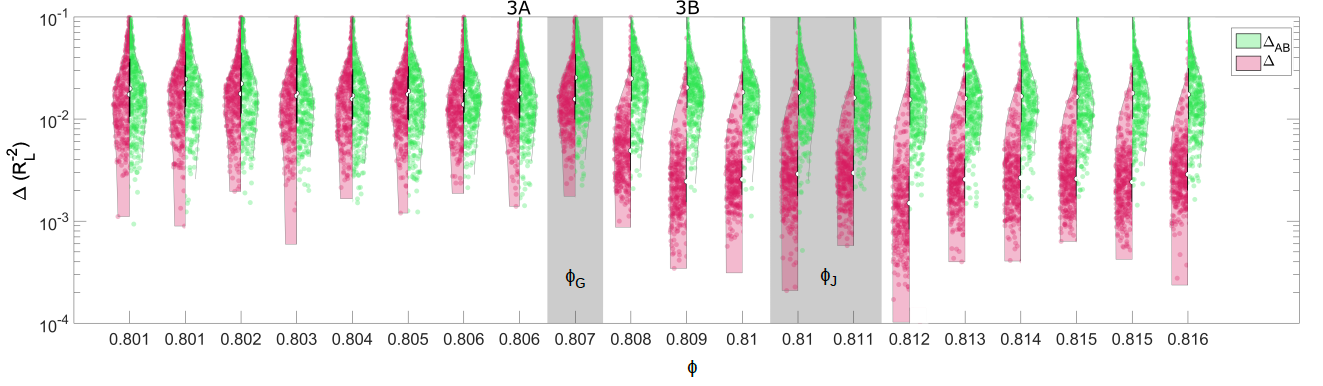}
    \caption{Side-by-side histograms of the probability density functions of $\Delta$ (magenta, left) and $\Delta_{AB}$ (green, right) measured at each density $\phi$. At $\phi < \phi_G$ (transition denoted by gray rectangle), the histograms of  $\Delta$ and $\Delta_{AB}$ agree (both in mean and width of the distribution), whereas for $\phi>\phi_G$, we see that $\Delta<\Delta_{AB}$. Above the jamming point $\phi_J$ (transition denoted by gray rectangle), the histograms of $\Delta$ and $\Delta_{AB}$ differ markedly. 
    Histograms 3A and 3B correspond to the two snapshots presented in Fig.~\ref{fig:Overlay}.}
    \label{fig:Violin}
\end{figure*}

\begin{figure}
    \includegraphics[width=0.9\linewidth]{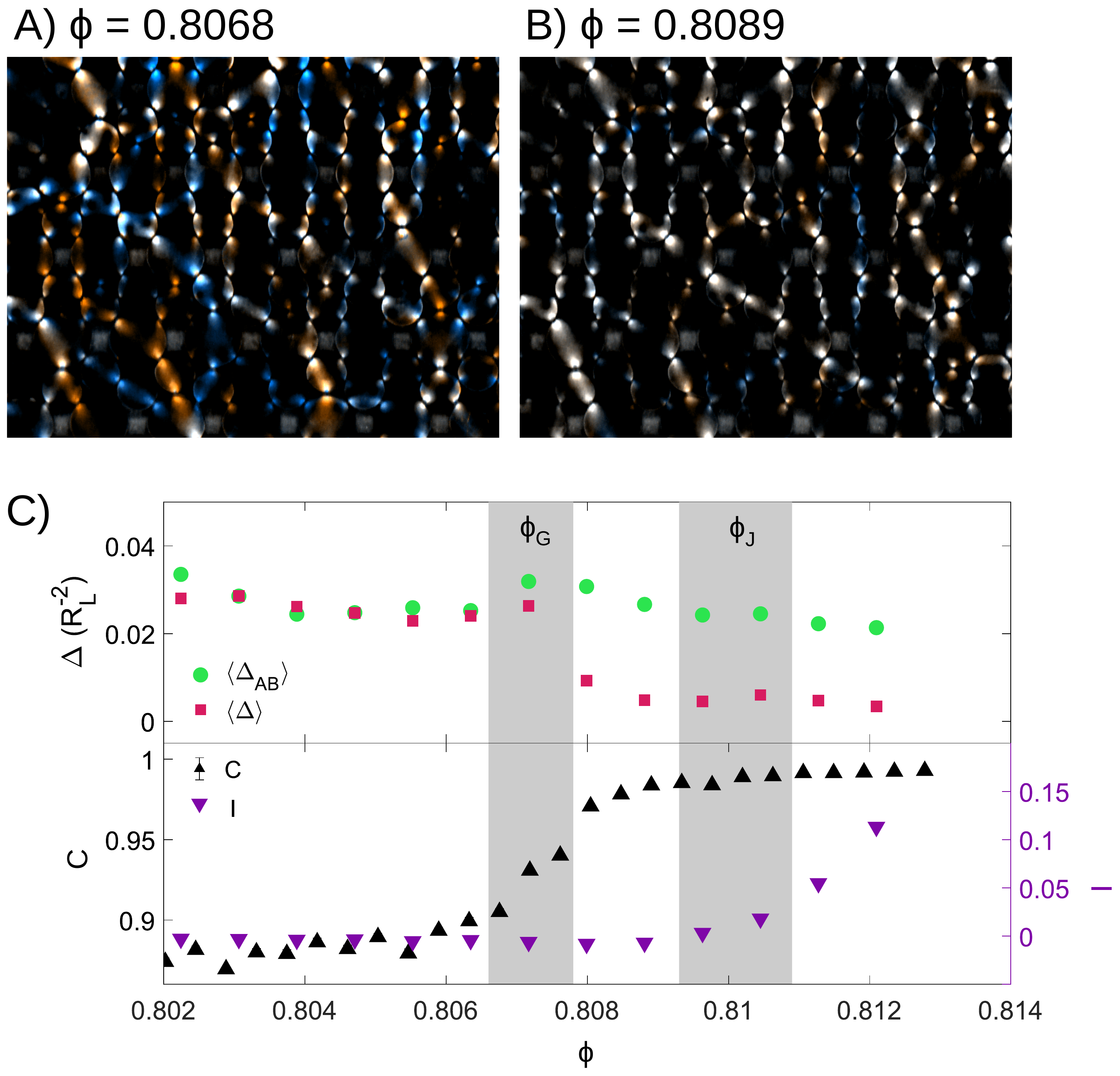}
    \caption{Overlay of two marginally stable states (red and blue) and their overlap (white) for (A) $\phi < \phi_G$ and (B) $\phi > \phi_G$. The two states have little overlap for $\phi < \phi_G$, whereas for $\phi>\phi_G$ the two states have a large overlap between their force network. A movie of overlays with increasing density makes this point even more saliently~\cite{footSI}. (C) Overlay of $\langle \Delta \rangle$ (magenta $\blacksquare$) and $\langle \Delta_{AB} \rangle$ (green $\bullet$) as a function of $\phi$ on the top axis; force correlation $C$ (black $\blacktriangle$) and fringe intensity $I$ (purple $\blacktriangledown$) on the bottom axis, both as function of $\phi$. 
    }
    \label{fig:Overlay}
\end{figure}

Figure~\ref{fig:Violin} presents the histograms of $\Delta$ and $\Delta_{AB}$ measured at various $\phi$. At low $\phi$, the statistical distributions of $\Delta$ and $\Delta_{AB}$ are nearly identical, which is characteristic of a normal solid. In contrast, for $\phi \gtrsim 0.807$ we observe that $\langle\Delta\rangle < \langle\Delta_{AB}\rangle$, indicating the onset of a Gardner-like regime at $\phi_G=0.807 \pm 0.0005$. As $\phi$ further increases, force chains emerge, thus identifying the jamming point, $\phi_J = 0.8100\pm 0.0005$~(Fig.~\ref{fig:Overlay}C); this value is determined by measuring the average proportional change  of the pixel intensity $I_g$ of the photoelastic (green) channel above the minimum observed value \cite{footSI}. Note that although we expect $\Delta=0$ in the jammed phase, a finite value is measured; this captures the noise floor of our system and analysis.


Having identified a Gardner-like transition using the particle displacement data, we now separately consider the evolution of inter-particle forces within each (marginally stable) state at different $\phi$. We observe changes in the persistence of the photoelastic fringes (as proxy for inter-particle forces, see Fig.~\ref{fig:Overlay}A-B), of a given state by compressing the system to a jammed reference density
$\phi_\mathrm{ref}= 0.8147 \pm 0.0002$ which is slightly above 
$\phi_\mathrm{max} = 0.8127 \pm 0.0002$, at which a better imaging is obtained (see \cite{footSI} for details).
For systems with few interparticle contacts, the correlation between fringes is dominated by noise, whereas in well-jammed systems the force network is completely percolated (due to the crystalline nature of the system), thus making  changes to the force network insignificant compared to the average inter-particle force.

Changes in the photo-elastic fringes at $\phi_\mathrm{ref}$ are  determined as follows. We first image the photoelastic fringes of the initial state, $\mathcal{I}$. The system is then decompressed to $\phi_\mathrm{min} < \phi< \phi_\mathrm{max}$, and evolved for $10 t_r$ (sufficient for the force network to randomize), before recompressing to $\phi_\mathrm{ref}$ and to image the photoelastic fringes of this final state, $\mathcal{F}$. We repeat this protocol for 30 equidistant densities within the interval $[\phi_\mathrm{min},\phi_\mathrm{max}]$. In all cases, the system is decompressed to $\phi_\mathrm{min}$ before moving to the next $\phi$ to erase any memory of the previous experiment~\cite{footSI}. We quantify the degree of similarity between the $\mathcal{I}$ and $\mathcal{F}$ states for a given $\phi$ using a normalized cross-correlation of the photoelastic fringes, taken $10 t_r$ apart:
\begin{equation}
    \label{eq:Force_similarity}
    C(\phi) = \Bigg \langle \frac{\sum_{x,y} [\mathcal{I}_{i} (x,y) - \overline{\mathcal{I}}_{i}][\mathcal{F}_{i} (x, y) - \overline{\mathcal{F}}_{i}]}{\sqrt{\sum_{x,y} [\mathcal{I}_{i}(x,y) - \overline{\mathcal{I}}_{i}]^2 \sum_{x,y} [\mathcal{F}_{i} (x,y) - \overline{\mathcal{F}}_{i}]^2}} \Bigg \rangle
\end{equation}
with $\mathcal{I}_i (x,y)$ the pixel intensity of pixel $(x,y)$ of particle $i$, $\overline{\mathcal{I}_i}$ the average pixel intensity of particle $i$ in state $\mathcal{I}$, and the average, $\langle\cdot\rangle$, running over all particles in all pairs $\mathcal{I}$ and $\mathcal{F}$ of states at a given $\phi$.

Figure~\ref{fig:Overlay} shows two superimposed images of force chains: $\mathcal{I}$ (blue) obtained before the airjet sweeps, and $\mathcal{F}$ (red) after, such that white denotes regions where force chains did not change, while red and blue denote force chains present in only one of the two images. For $\phi< \phi_G$ (Fig.~\ref{fig:Overlay}A) the rare force chain overlaps (white) indicate that inter-particle contacts remain variable at low $\phi$. In contrast,  for $\phi > \phi_G$ (Fig.~\ref{fig:Overlay}B) white regions dominate,  indicating that inter-particle contacts persist. Similar images obtained over the full density range further reveal that force-chain rearrangements are long-range, even though the particle rearrangements are not~\cite{footSI}. 

Figure~\ref{fig:Overlay}C compares these perspectives, showing $\langle\Delta\rangle$ and $\langle\Delta_{AB}\rangle$, the measure for inter-particle forces $\langle I_{g}\rangle$, and the normalized cross-correlation $C$, all as a function of $\phi$. This comparison makes it clear that the onset of the Gardner-like regime (for which $\langle\Delta\rangle < \langle\Delta_{AB}]\rangle$) coincides with the onset of the conservation of inter-particle contacts (given by the sharp rise in $C$), and is distinct from $\phi_J$ (defined in the onset of the rise of $\langle I_{g}\rangle$), suggesting that the force network gets increasingly determined as soon as $\phi>\phi_G$, which is well before $\phi_J$.

\begin{figure}
    \includegraphics[width=\linewidth]{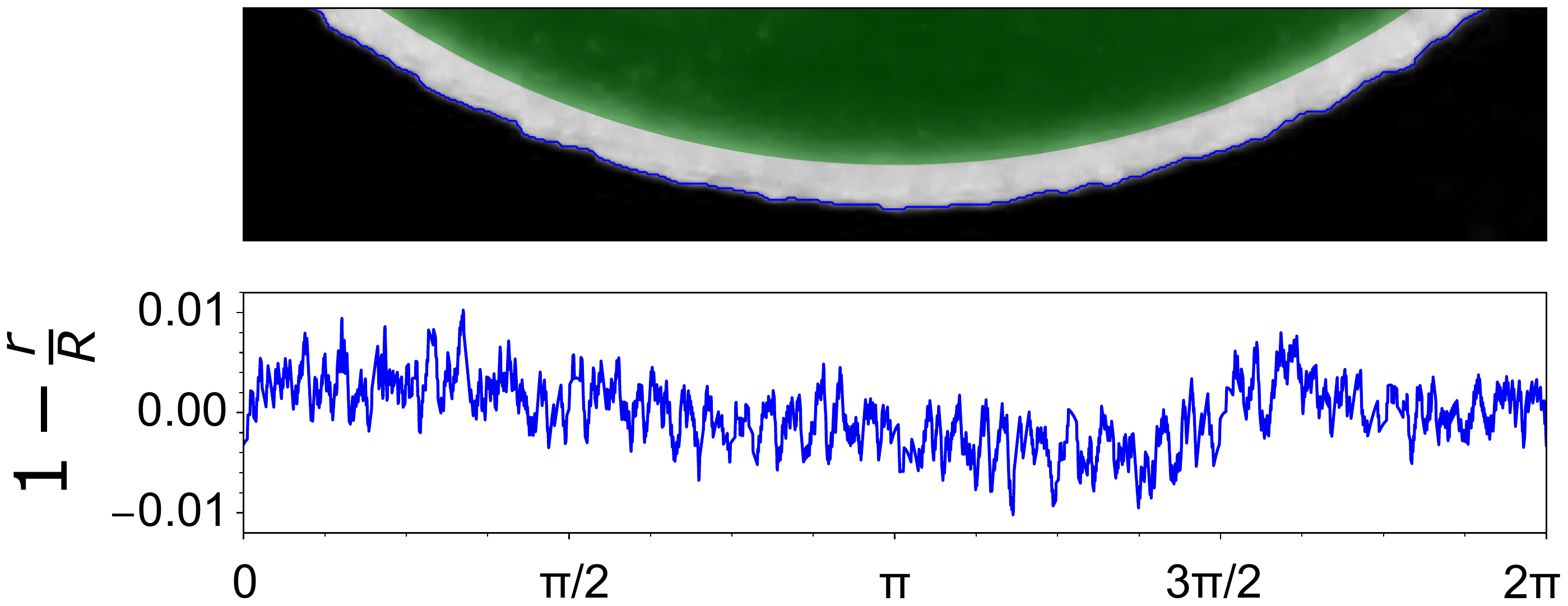}
    \caption{(a) Segment of a micrograph of a single photoelastic particle, with edge-detection shown in blue; the green circle traces a perfect circle for reference. (b) Fractional deviation of the particle edge $r$ from the average radius $R$ along the circumference.}
    \label{fig:roughness}
\end{figure}

By analogy to what has been reported for numerical simulations of size-polydisperse particles in otherwise crystalline systems \cite{charbonneau2019glassy}, we expect the distance to jamming to be controlled by particle disorder. Given that all particles were cut from flat sheets with the same fixed-radius metal cutter \cite{daniels2017photoelastic}, disorder is here expected to be dominated by irregularities along the disk edges (see Fig.~\ref{fig:roughness}). Generalizing the polydispersity argument of \cite{charbonneau2019glassy} to this case, we expect the onset of the Gardner regime to be set by the particles' dimensionless deviation from a constant radius: 
\begin{equation}
    \sqrt{\phi_J - \phi_G} \propto 1 - \frac{r}{R}.
    \label{eq:roughness}
\end{equation}
Two key features emerge from the image analysis: asphericity of $\sim 1\%$, superimposed with a surface roughness of $\sim 0.3\%$. Remarkably, both quantities are of the same order of magnitude as the relative distance between the Gardner-like transition and the jamming point for our system, $s = \sqrt{(0.810 - 0.807)/0.81} \approx 1\%$.

\emph{Conclusion --} We have shown that small particle irregularities---always present in experiments but often neglected---in an otherwise crystalline system exhibit Gardner-physics-like features near jamming. Although 2D systems are expected not to exhibit proper Gardner criticality~\cite{berthier2019gardner}, the finite size of our system suppresses the long wavelength fluctuations that would normally occlude  this effect in the thermodynamic limit, thus preserving some of its physical features. This choice of system further allowed us to study changes in the force network of one specific configuration, where we found experimental evidence correlating the onset of that regime to the determination of force contacts near jamming, and relate the distance to jamming of that transition to the inherent roughness of the particles.

This work motivates delving further into the influence of particle roughness on a micromechanical level. Whereas surface roughness has been investigated on a macroscopic level, such as for relating surface roughness and friction in glass spheres  \cite[\textit{e.g.}]{utermann2011tailoring}, we showed that roughness also changes interactions on a microscopic level. This finding raises questions about what signatures of frictional jamming found for smooth particles \cite{silbert2002geometry} match those needed to describe, more realistic, rough particles \cite{pradeep2021jamming}. In this context, including geometrical asperities \cite{papanikolaou2013isostaticity} in numerical simulations could provide particularly invaluable insight. 

\begin{acknowledgments}
    \paragraph{Acknowledgements} We thank Sho Yaida, Jonathan Kollmer, and Eric Corwin for motivating discussions, Joshua Dijksman for help with image processing, and Jonathan Kollmer, Clayton Kirberger, and Josh Miller for technical assistance. PC acknowledges support from the Simons Foundation, Grant No. 454937. KD acknowledges support from NSF DMR-0644743 (apparatus) and current support from DMR-2104986. LK acknowledges support from the European Union's Horizon 2020 research and innovation programme {\it Caliper},  Marie Sk{\l}odowska Curie Grant  No. 812638.

\end{acknowledgments}


\begin{thebibliography}{34}
\expandafter\ifx\csname natexlab\endcsname\relax\def\natexlab#1{#1}\fi
\expandafter\ifx\csname bibnamefont\endcsname\relax
  \def\bibnamefont#1{#1}\fi
\expandafter\ifx\csname bibfnamefont\endcsname\relax
  \def\bibfnamefont#1{#1}\fi
\expandafter\ifx\csname citenamefont\endcsname\relax
  \def\citenamefont#1{#1}\fi
\expandafter\ifx\csname url\endcsname\relax
  \def\url#1{\texttt{#1}}\fi
\expandafter\ifx\csname urlprefix\endcsname\relax\def\urlprefix{URL }\fi
\providecommand{\bibinfo}[2]{#2}
\providecommand{\eprint}[2][]{\url{#2}}

\bibitem[{\citenamefont{Dantu}(1968)}]{dantu1968etude}
\bibinfo{author}{\bibfnamefont{P.}~\bibnamefont{Dantu}},
  \bibinfo{journal}{G{\'e}otechnique} \textbf{\bibinfo{volume}{18}},
  \bibinfo{pages}{50} (\bibinfo{year}{1968}).

\bibitem[{\citenamefont{Jaeger et~al.}(1996)\citenamefont{Jaeger, Nagel, and
  Behringer}}]{jaeger1996granular}
\bibinfo{author}{\bibfnamefont{H.~M.} \bibnamefont{Jaeger}},
  \bibinfo{author}{\bibfnamefont{S.~R.} \bibnamefont{Nagel}}, \bibnamefont{and}
  \bibinfo{author}{\bibfnamefont{R.~P.} \bibnamefont{Behringer}},
  \bibinfo{journal}{Reviews of Modern Physics} \textbf{\bibinfo{volume}{68}},
  \bibinfo{pages}{1259} (\bibinfo{year}{1996}).

\bibitem[{\citenamefont{Radjai et~al.}(1998)\citenamefont{Radjai, Wolf, Jean,
  and Moreau}}]{radjai1998bimodal}
\bibinfo{author}{\bibfnamefont{F.}~\bibnamefont{Radjai}},
  \bibinfo{author}{\bibfnamefont{D.~E.} \bibnamefont{Wolf}},
  \bibinfo{author}{\bibfnamefont{M.}~\bibnamefont{Jean}}, \bibnamefont{and}
  \bibinfo{author}{\bibfnamefont{J.-J.} \bibnamefont{Moreau}},
  \bibinfo{journal}{Physical Review Letters} \textbf{\bibinfo{volume}{80}},
  \bibinfo{pages}{61} (\bibinfo{year}{1998}).

\bibitem[{\citenamefont{Howell et~al.}(1999)\citenamefont{Howell, Behringer,
  and Veje}}]{howell1999stress}
\bibinfo{author}{\bibfnamefont{D.}~\bibnamefont{Howell}},
  \bibinfo{author}{\bibfnamefont{R.~P.} \bibnamefont{Behringer}},
  \bibnamefont{and} \bibinfo{author}{\bibfnamefont{C.}~\bibnamefont{Veje}},
  \bibinfo{journal}{Physical Review Letters} \textbf{\bibinfo{volume}{82}},
  \bibinfo{pages}{5241} (\bibinfo{year}{1999}).

\bibitem[{\citenamefont{Peters et~al.}(2005)\citenamefont{Peters, Muthuswamy,
  Wibowo, and Tordesillas}}]{peters2005characterization}
\bibinfo{author}{\bibfnamefont{J.}~\bibnamefont{Peters}},
  \bibinfo{author}{\bibfnamefont{M.}~\bibnamefont{Muthuswamy}},
  \bibinfo{author}{\bibfnamefont{J.}~\bibnamefont{Wibowo}}, \bibnamefont{and}
  \bibinfo{author}{\bibfnamefont{A.}~\bibnamefont{Tordesillas}},
  \bibinfo{journal}{Physical Review E} \textbf{\bibinfo{volume}{72}},
  \bibinfo{pages}{041307} (\bibinfo{year}{2005}).

\bibitem[{\citenamefont{Tighe et~al.}(2010)\citenamefont{Tighe, Snoeijer,
  Vlugt, and van Hecke}}]{tighe2010force}
\bibinfo{author}{\bibfnamefont{B.~P.} \bibnamefont{Tighe}},
  \bibinfo{author}{\bibfnamefont{J.~H.} \bibnamefont{Snoeijer}},
  \bibinfo{author}{\bibfnamefont{T.~J.} \bibnamefont{Vlugt}}, \bibnamefont{and}
  \bibinfo{author}{\bibfnamefont{M.}~\bibnamefont{van Hecke}},
  \bibinfo{journal}{Soft Matter} \textbf{\bibinfo{volume}{6}},
  \bibinfo{pages}{2908} (\bibinfo{year}{2010}).

\bibitem[{\citenamefont{Kollmer and Daniels}(2019)}]{kollmer2019betweenness}
\bibinfo{author}{\bibfnamefont{J.~E.} \bibnamefont{Kollmer}} \bibnamefont{and}
  \bibinfo{author}{\bibfnamefont{K.~E.} \bibnamefont{Daniels}},
  \bibinfo{journal}{Soft Matter} \textbf{\bibinfo{volume}{15}},
  \bibinfo{pages}{1793} (\bibinfo{year}{2019}).

\bibitem[{\citenamefont{Charbonneau et~al.}(2017)\citenamefont{Charbonneau,
  Kurchan, Parisi, Urbani, and Zamponi}}]{charbonneau2017glass}
\bibinfo{author}{\bibfnamefont{P.}~\bibnamefont{Charbonneau}},
  \bibinfo{author}{\bibfnamefont{J.}~\bibnamefont{Kurchan}},
  \bibinfo{author}{\bibfnamefont{G.}~\bibnamefont{Parisi}},
  \bibinfo{author}{\bibfnamefont{P.}~\bibnamefont{Urbani}}, \bibnamefont{and}
  \bibinfo{author}{\bibfnamefont{F.}~\bibnamefont{Zamponi}},
  \bibinfo{journal}{Annual Review of Condensed Matter Physics}
  \textbf{\bibinfo{volume}{8}}, \bibinfo{pages}{265} (\bibinfo{year}{2017}).

\bibitem[{\citenamefont{Charbonneau and Morse}(2021)}]{charbonneau2021memory}
\bibinfo{author}{\bibfnamefont{P.}~\bibnamefont{Charbonneau}} \bibnamefont{and}
  \bibinfo{author}{\bibfnamefont{P.~K.} \bibnamefont{Morse}},
  \bibinfo{journal}{Physical Review Letters} \textbf{\bibinfo{volume}{126}},
  \bibinfo{pages}{088001} (\bibinfo{year}{2021}).

\bibitem[{\citenamefont{Gardner}(1985)}]{gardner1985spin}
\bibinfo{author}{\bibfnamefont{E.}~\bibnamefont{Gardner}},
  \bibinfo{journal}{Nuclear Physics B} \textbf{\bibinfo{volume}{257}},
  \bibinfo{pages}{747} (\bibinfo{year}{1985}).

\bibitem[{\citenamefont{Kurchan et~al.}(2013)\citenamefont{Kurchan, Parisi,
  Urbani, and Zamponi}}]{kurchan2013exact}
\bibinfo{author}{\bibfnamefont{J.}~\bibnamefont{Kurchan}},
  \bibinfo{author}{\bibfnamefont{G.}~\bibnamefont{Parisi}},
  \bibinfo{author}{\bibfnamefont{P.}~\bibnamefont{Urbani}}, \bibnamefont{and}
  \bibinfo{author}{\bibfnamefont{F.}~\bibnamefont{Zamponi}},
  \bibinfo{journal}{The Journal of Physical Chemistry B}
  \textbf{\bibinfo{volume}{117}}, \bibinfo{pages}{12979}
  (\bibinfo{year}{2013}).

\bibitem[{\citenamefont{Charbonneau et~al.}(2019)\citenamefont{Charbonneau,
  Corwin, Fu, Tsekenis, and van Der~Naald}}]{charbonneau2019glassy}
\bibinfo{author}{\bibfnamefont{P.}~\bibnamefont{Charbonneau}},
  \bibinfo{author}{\bibfnamefont{E.~I.} \bibnamefont{Corwin}},
  \bibinfo{author}{\bibfnamefont{L.}~\bibnamefont{Fu}},
  \bibinfo{author}{\bibfnamefont{G.}~\bibnamefont{Tsekenis}}, \bibnamefont{and}
  \bibinfo{author}{\bibfnamefont{M.}~\bibnamefont{van Der~Naald}},
  \bibinfo{journal}{Physical Review E} \textbf{\bibinfo{volume}{99}},
  \bibinfo{pages}{020901} (\bibinfo{year}{2019}).

\bibitem[{\citenamefont{Berthier et~al.}(2019)\citenamefont{Berthier, Biroli,
  Charbonneau, Corwin, Franz, and Zamponi}}]{berthier2019gardner}
\bibinfo{author}{\bibfnamefont{L.}~\bibnamefont{Berthier}},
  \bibinfo{author}{\bibfnamefont{G.}~\bibnamefont{Biroli}},
  \bibinfo{author}{\bibfnamefont{P.}~\bibnamefont{Charbonneau}},
  \bibinfo{author}{\bibfnamefont{E.~I.} \bibnamefont{Corwin}},
  \bibinfo{author}{\bibfnamefont{S.}~\bibnamefont{Franz}}, \bibnamefont{and}
  \bibinfo{author}{\bibfnamefont{F.}~\bibnamefont{Zamponi}},
  \bibinfo{journal}{The Journal of Chemical Physics}
  \textbf{\bibinfo{volume}{151}}, \bibinfo{pages}{010901}
  (\bibinfo{year}{2019}).

\bibitem[{\citenamefont{Charbonneau et~al.}(2014)\citenamefont{Charbonneau,
  Kurchan, Parisi, Urbani, and Zamponi}}]{charbonneau2014fractal}
\bibinfo{author}{\bibfnamefont{P.}~\bibnamefont{Charbonneau}},
  \bibinfo{author}{\bibfnamefont{J.}~\bibnamefont{Kurchan}},
  \bibinfo{author}{\bibfnamefont{G.}~\bibnamefont{Parisi}},
  \bibinfo{author}{\bibfnamefont{P.}~\bibnamefont{Urbani}}, \bibnamefont{and}
  \bibinfo{author}{\bibfnamefont{F.}~\bibnamefont{Zamponi}},
  \bibinfo{journal}{Nature Communications} \textbf{\bibinfo{volume}{5}},
  \bibinfo{pages}{3725} (\bibinfo{year}{2014}).

\bibitem[{\citenamefont{Tsekenis}(2021)}]{tsekenis2021jamming}
\bibinfo{author}{\bibfnamefont{G.}~\bibnamefont{Tsekenis}},
  \bibinfo{journal}{EPL (Europhysics Letters)} \textbf{\bibinfo{volume}{135}},
  \bibinfo{pages}{36001} (\bibinfo{year}{2021}).

\bibitem[{\citenamefont{Hammond and Corwin}(2020)}]{hammond2020experimental}
\bibinfo{author}{\bibfnamefont{A.~P.} \bibnamefont{Hammond}} \bibnamefont{and}
  \bibinfo{author}{\bibfnamefont{E.~I.} \bibnamefont{Corwin}},
  \bibinfo{journal}{Proceedings of the National Academy of Sciences}
  \textbf{\bibinfo{volume}{117}}, \bibinfo{pages}{5714} (\bibinfo{year}{2020}).

\bibitem[{\citenamefont{Biroli and Urbani}(2016)}]{biroli2016breakdown}
\bibinfo{author}{\bibfnamefont{G.}~\bibnamefont{Biroli}} \bibnamefont{and}
  \bibinfo{author}{\bibfnamefont{P.}~\bibnamefont{Urbani}},
  \bibinfo{journal}{Nature Physics} \textbf{\bibinfo{volume}{12}},
  \bibinfo{pages}{1130} (\bibinfo{year}{2016}).

\bibitem[{\citenamefont{Jin and Yoshino}(2017)}]{jin2017exploring}
\bibinfo{author}{\bibfnamefont{Y.}~\bibnamefont{Jin}} \bibnamefont{and}
  \bibinfo{author}{\bibfnamefont{H.}~\bibnamefont{Yoshino}},
  \bibinfo{journal}{Nature Communications} \textbf{\bibinfo{volume}{8}},
  \bibinfo{pages}{1} (\bibinfo{year}{2017}).

\bibitem[{\citenamefont{Seguin and Dauchot}(2016)}]{seguin2016experimental}
\bibinfo{author}{\bibfnamefont{A.}~\bibnamefont{Seguin}} \bibnamefont{and}
  \bibinfo{author}{\bibfnamefont{O.}~\bibnamefont{Dauchot}},
  \bibinfo{journal}{Physical Review Letters} \textbf{\bibinfo{volume}{117}},
  \bibinfo{pages}{228001} (\bibinfo{year}{2016}).

\bibitem[{\citenamefont{Xiao et~al.}(2021)\citenamefont{Xiao, Liu, and
  Durian}}]{xiao2021probing}
\bibinfo{author}{\bibfnamefont{H.}~\bibnamefont{Xiao}},
  \bibinfo{author}{\bibfnamefont{A.~J.} \bibnamefont{Liu}}, \bibnamefont{and}
  \bibinfo{author}{\bibfnamefont{D.~J.} \bibnamefont{Durian}},
  \bibinfo{journal}{arXiv preprint arXiv:2111.03921}  (\bibinfo{year}{2021}).

\bibitem[{\citenamefont{Likos and Henley}(1993)}]{likos1993complex}
\bibinfo{author}{\bibfnamefont{C.}~\bibnamefont{Likos}} \bibnamefont{and}
  \bibinfo{author}{\bibfnamefont{C.}~\bibnamefont{Henley}},
  \bibinfo{journal}{Philosophical Magazine B} \textbf{\bibinfo{volume}{68}},
  \bibinfo{pages}{85} (\bibinfo{year}{1993}).

\bibitem[{\citenamefont{Berthier et~al.}(2016)\citenamefont{Berthier,
  Charbonneau, Jin, Parisi, Seoane, and Zamponi}}]{berthier2016growing}
\bibinfo{author}{\bibfnamefont{L.}~\bibnamefont{Berthier}},
  \bibinfo{author}{\bibfnamefont{P.}~\bibnamefont{Charbonneau}},
  \bibinfo{author}{\bibfnamefont{Y.}~\bibnamefont{Jin}},
  \bibinfo{author}{\bibfnamefont{G.}~\bibnamefont{Parisi}},
  \bibinfo{author}{\bibfnamefont{B.}~\bibnamefont{Seoane}}, \bibnamefont{and}
  \bibinfo{author}{\bibfnamefont{F.}~\bibnamefont{Zamponi}},
  \bibinfo{journal}{Proceedings of the National Academy of Sciences}
  \textbf{\bibinfo{volume}{113}}, \bibinfo{pages}{8397} (\bibinfo{year}{2016}).

\bibitem[{\citenamefont{Lerner et~al.}(2013)\citenamefont{Lerner, D{\"u}ring,
  and Wyart}}]{lerner2013low}
\bibinfo{author}{\bibfnamefont{E.}~\bibnamefont{Lerner}},
  \bibinfo{author}{\bibfnamefont{G.}~\bibnamefont{D{\"u}ring}},
  \bibnamefont{and} \bibinfo{author}{\bibfnamefont{M.}~\bibnamefont{Wyart}},
  \bibinfo{journal}{Soft Matter} \textbf{\bibinfo{volume}{9}},
  \bibinfo{pages}{8252} (\bibinfo{year}{2013}).

\bibitem[{\citenamefont{Charbonneau
  et~al.}(2015{\natexlab{a}})\citenamefont{Charbonneau, Jin, Parisi, Rainone,
  Seoane, and Zamponi}}]{charbonneau2015numerical}
\bibinfo{author}{\bibfnamefont{P.}~\bibnamefont{Charbonneau}},
  \bibinfo{author}{\bibfnamefont{Y.}~\bibnamefont{Jin}},
  \bibinfo{author}{\bibfnamefont{G.}~\bibnamefont{Parisi}},
  \bibinfo{author}{\bibfnamefont{C.}~\bibnamefont{Rainone}},
  \bibinfo{author}{\bibfnamefont{B.}~\bibnamefont{Seoane}}, \bibnamefont{and}
  \bibinfo{author}{\bibfnamefont{F.}~\bibnamefont{Zamponi}},
  \bibinfo{journal}{Physical Review E} \textbf{\bibinfo{volume}{92}},
  \bibinfo{pages}{012316} (\bibinfo{year}{2015}{\natexlab{a}}).

\bibitem[{\citenamefont{Charbonneau
  et~al.}(2015{\natexlab{b}})\citenamefont{Charbonneau, Corwin, Parisi, and
  Zamponi}}]{charbonneau2015jamming}
\bibinfo{author}{\bibfnamefont{P.}~\bibnamefont{Charbonneau}},
  \bibinfo{author}{\bibfnamefont{E.~I.} \bibnamefont{Corwin}},
  \bibinfo{author}{\bibfnamefont{G.}~\bibnamefont{Parisi}}, \bibnamefont{and}
  \bibinfo{author}{\bibfnamefont{F.}~\bibnamefont{Zamponi}},
  \bibinfo{journal}{Physical Review Letters} \textbf{\bibinfo{volume}{114}},
  \bibinfo{pages}{125504} (\bibinfo{year}{2015}{\natexlab{b}}).

\bibitem[{\citenamefont{Puckett and Daniels}(2013)}]{puckett2013equilibrating}
\bibinfo{author}{\bibfnamefont{J.~G.} \bibnamefont{Puckett}} \bibnamefont{and}
  \bibinfo{author}{\bibfnamefont{K.~E.} \bibnamefont{Daniels}},
  \bibinfo{journal}{Physical Review Letters} \textbf{\bibinfo{volume}{110}},
  \bibinfo{pages}{058001} (\bibinfo{year}{2013}).

\bibitem[{\citenamefont{Bililign et~al.}(2019)\citenamefont{Bililign, Kollmer,
  and Daniels}}]{bililign2019protocol}
\bibinfo{author}{\bibfnamefont{E.~S.} \bibnamefont{Bililign}},
  \bibinfo{author}{\bibfnamefont{J.~E.} \bibnamefont{Kollmer}},
  \bibnamefont{and} \bibinfo{author}{\bibfnamefont{K.~E.}
  \bibnamefont{Daniels}}, \bibinfo{journal}{Physical Review Letters}
  \textbf{\bibinfo{volume}{122}}, \bibinfo{pages}{038001}
  (\bibinfo{year}{2019}).

\bibitem[{\citenamefont{Kollmer}()}]{PeGS}
\bibinfo{author}{\bibfnamefont{J.}~\bibnamefont{Kollmer}},
  \emph{\bibinfo{title}{Photo-elastic grain solver}},
  \urlprefix\url{https://github.com/jekollmer/PEGS}.

\bibitem[{\citenamefont{Daniels et~al.}(2017)\citenamefont{Daniels, Kollmer,
  and Puckett}}]{daniels2017photoelastic}
\bibinfo{author}{\bibfnamefont{K.~E.} \bibnamefont{Daniels}},
  \bibinfo{author}{\bibfnamefont{J.~E.} \bibnamefont{Kollmer}},
  \bibnamefont{and} \bibinfo{author}{\bibfnamefont{J.~G.}
  \bibnamefont{Puckett}}, \bibinfo{journal}{Review of Scientific Instruments}
  \textbf{\bibinfo{volume}{88}}, \bibinfo{pages}{051808}
  (\bibinfo{year}{2017}).

\bibitem[{foo()}]{footSI}
\bibinfo{note}{See Supplementary Material for a schematic overview of the
  protocols, details about the MSD correction and orthogonalization schemes,
  the jamming point determination, and a movie of force contact overlaps.}

\bibitem[{\citenamefont{Utermann et~al.}(2011)\citenamefont{Utermann, Aurin,
  Benderoth, Fischer, and Schr{\"o}ter}}]{utermann2011tailoring}
\bibinfo{author}{\bibfnamefont{S.}~\bibnamefont{Utermann}},
  \bibinfo{author}{\bibfnamefont{P.}~\bibnamefont{Aurin}},
  \bibinfo{author}{\bibfnamefont{M.}~\bibnamefont{Benderoth}},
  \bibinfo{author}{\bibfnamefont{C.}~\bibnamefont{Fischer}}, \bibnamefont{and}
  \bibinfo{author}{\bibfnamefont{M.}~\bibnamefont{Schr{\"o}ter}},
  \bibinfo{journal}{Physical Review E} \textbf{\bibinfo{volume}{84}},
  \bibinfo{pages}{031306} (\bibinfo{year}{2011}).

\bibitem[{\citenamefont{Silbert et~al.}(2002)\citenamefont{Silbert,
  Erta{\c{s}}, Grest, Halsey, and Levine}}]{silbert2002geometry}
\bibinfo{author}{\bibfnamefont{L.~E.} \bibnamefont{Silbert}},
  \bibinfo{author}{\bibfnamefont{D.}~\bibnamefont{Erta{\c{s}}}},
  \bibinfo{author}{\bibfnamefont{G.~S.} \bibnamefont{Grest}},
  \bibinfo{author}{\bibfnamefont{T.~C.} \bibnamefont{Halsey}},
  \bibnamefont{and} \bibinfo{author}{\bibfnamefont{D.}~\bibnamefont{Levine}},
  \bibinfo{journal}{Physical Review E} \textbf{\bibinfo{volume}{65}},
  \bibinfo{pages}{031304} (\bibinfo{year}{2002}).

\bibitem[{\citenamefont{Pradeep et~al.}(2021)\citenamefont{Pradeep, Nabizadeh,
  Jacob, Jamali, and Hsiao}}]{pradeep2021jamming}
\bibinfo{author}{\bibfnamefont{S.}~\bibnamefont{Pradeep}},
  \bibinfo{author}{\bibfnamefont{M.}~\bibnamefont{Nabizadeh}},
  \bibinfo{author}{\bibfnamefont{A.~R.} \bibnamefont{Jacob}},
  \bibinfo{author}{\bibfnamefont{S.}~\bibnamefont{Jamali}}, \bibnamefont{and}
  \bibinfo{author}{\bibfnamefont{L.~C.} \bibnamefont{Hsiao}},
  \bibinfo{journal}{Physical Review Letters} \textbf{\bibinfo{volume}{127}},
  \bibinfo{pages}{158002} (\bibinfo{year}{2021}).

\bibitem[{\citenamefont{Papanikolaou et~al.}(2013)\citenamefont{Papanikolaou,
  O’Hern, and Shattuck}}]{papanikolaou2013isostaticity}
\bibinfo{author}{\bibfnamefont{S.}~\bibnamefont{Papanikolaou}},
  \bibinfo{author}{\bibfnamefont{C.~S.} \bibnamefont{O’Hern}},
  \bibnamefont{and} \bibinfo{author}{\bibfnamefont{M.~D.}
  \bibnamefont{Shattuck}}, \bibinfo{journal}{Physical Review Letters}
  \textbf{\bibinfo{volume}{110}}, \bibinfo{pages}{198002}
  (\bibinfo{year}{2013}).

\end{thebibliography}

\clearpage
\onecolumngrid
\appendix

\setcounter{equation}{0}
\renewcommand{\thefigure}{A.\arabic{figure}}
\renewcommand{\theequation}{A.\arabic{equation}}

\centerline{\bf \large Gardner-like transition from variable to persistent force contacts in granular crystals}

\section*{Supplementary Information}

\centerline{Lars Kool,$^{1,2}$ Patrick Charbonneau,$^3$ and Karen E. Daniels$^2$}

\bigskip

\centerline{$^1${\it Laboratoire de Physique et M\'echanique des Milieux H\'et\`erog\`enes, ESPCI, Paris, France}}

\centerline{$^2${\it Department of Physics, North Carolina State University, Raleigh, North Carolina 27695, USA}}

\centerline{$^3${\it Departments of Chemistry \& Physics, Duke University, Durham, North Carolina 27708, USA}}

\bigskip 

\twocolumngrid 

\subsection{Description of Movie}

The movie (available online) provides a fuller display of the transition from Fig.~\ref{fig:Overlay}A to B.  Each frame overlays two experimental snapshots: one  taken before (blue) and one taken after (red) a single airjet sweep, obtained for gradually increasing $\phi$. The overlap between the two images appears as white. For $\phi < 0.807$,  sparse overlaps (lack of white) of the force chains is clear; far above  $\phi_G > 0.807$, the overlap drastically increases (white becomes dominant). The quantification of the  transition from variable to persistent force contacts is reported in Fig.~\ref{fig:Overlay}C.

\subsection{Determining the jamming transition}

We measure the jamming transition $\phi_J$ using the photoelastic stresses, visualized using circularly polarized green light. At the onset of jamming, fringes start to develop, while other  contributions to the  green signal in the image are minimal and constant (\textit{e.g.}, scattering of green light off the porous grid under the particles). We  therefore normalize the average green light intensity $I_g$ according to
\begin{equation*}
I(\phi) = \frac{\langle I_g(\phi) \rangle}{\langle I_g(\phi_{min}) \rangle}
\end{equation*}
where $\langle I_g(\phi_\mathrm{min}) \rangle$ is the background green intensity, measured at the lowest $\phi$ and $\langle\cdot\rangle$ denotes the average taken over all pixels at least 2$d_l$ from the boundaries and all images taken at that $\phi$.  

At low $\phi$, there are no photoelastic fringes present in the image ($I_g$ stays within 0.5\% of $I_g(\phi_\mathrm{min})$), indicating that the system is not yet jammed. For $\phi > 0.8096$, $I_g$ lies above the background level, but remains within the noise floor of 0.5\%; only at $\phi = 0.8105$ (greyed region in Fig.~\ref{fig:Overlay}C), does $I$ deviate significantly (1.8\%), thus indicating that the system has entered the jammed phase. We therefore identify the jamming transition as $\phi_J = 0.8100 \pm 0.0005)$.

\subsection{Mean squared displacement correction}

\begin{figure}
    \includegraphics[width=0.8\linewidth]{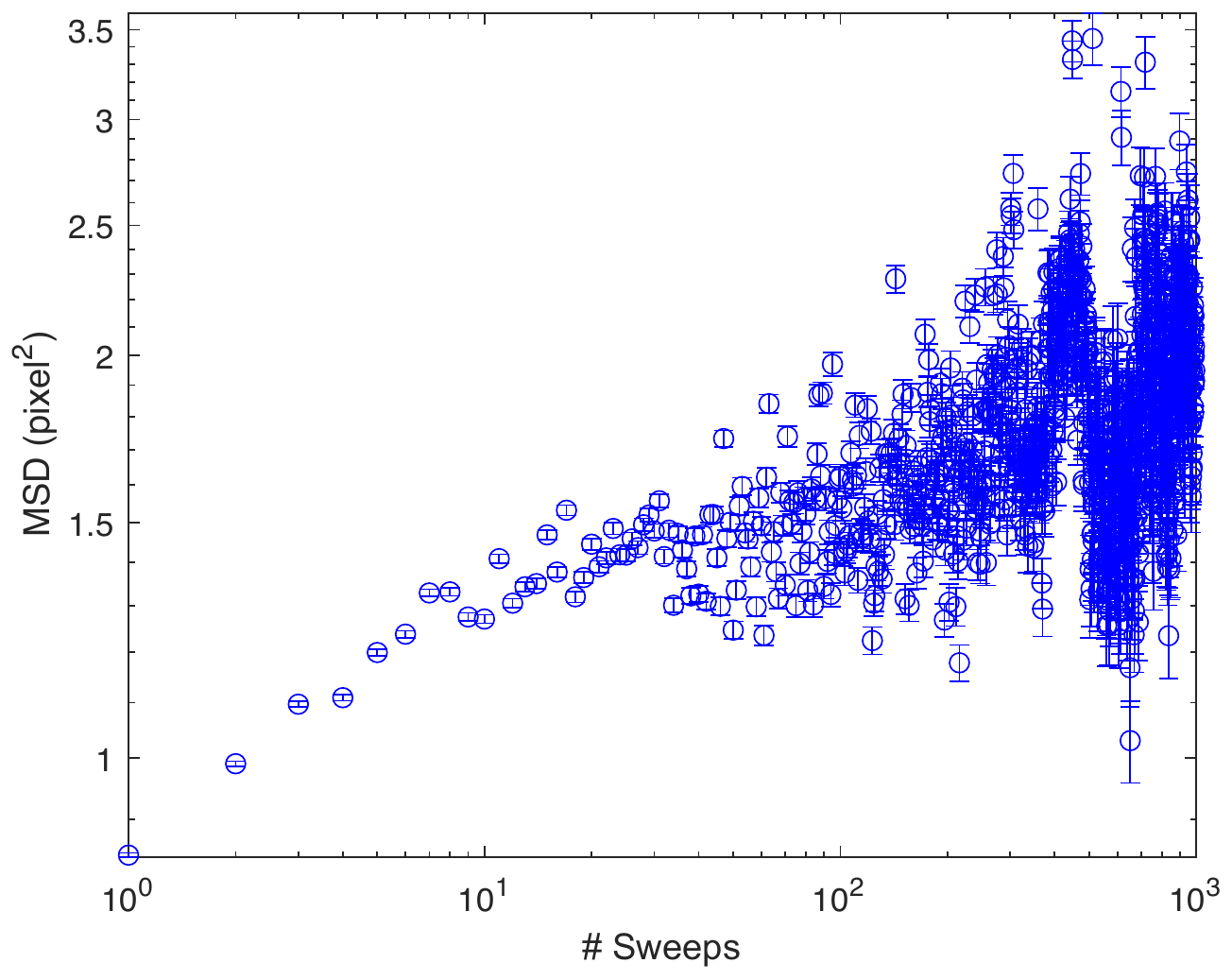}
    \caption{Time evolution of the MSD, showing plateau above a few hundred sweep cycles ($t_r$).}
    \label{fig:MSD_correction}
\end{figure}

In our experiments, we compare the mean cage size, $\Delta$, with the mean cage separation, $\Delta_{AB}$. However, due to the slow nature of the airjet sweeps used to randomize particle positions within their cage, the true long-time limit of MSD is not experimentally attainable. The full extent of our data collection, out to $1000 t_r$, is shown in Fig.~\ref{fig:MSD_correction}. 

We therefore determined the ratio between the long-time MSD (at $1000 t_r$) and an attainable timescale, $100 t_r$, and corrected all reported $\Delta$ values accordingly through a constant rescaling factor,
\begin{equation*}
D = \frac{\mathrm{MSD}(950)}{\mathrm{MSD}(100)} = 1.2
\end{equation*}
with $\mathrm{MSD}(950)$ the average MSD between 900$t_r$ and 1000$t_r$, and $\mathrm{MSD}(100)$ the average MSD between 95$t_r$ and 105$t_r$. We average over a range of MSD values, due to the difficulty of obtaining stochastically independent trajectories at longer timescales.

\newpage
\subsection{Schematic overview of the protocols}

\begin{figure*}
    \includegraphics[width=\linewidth]{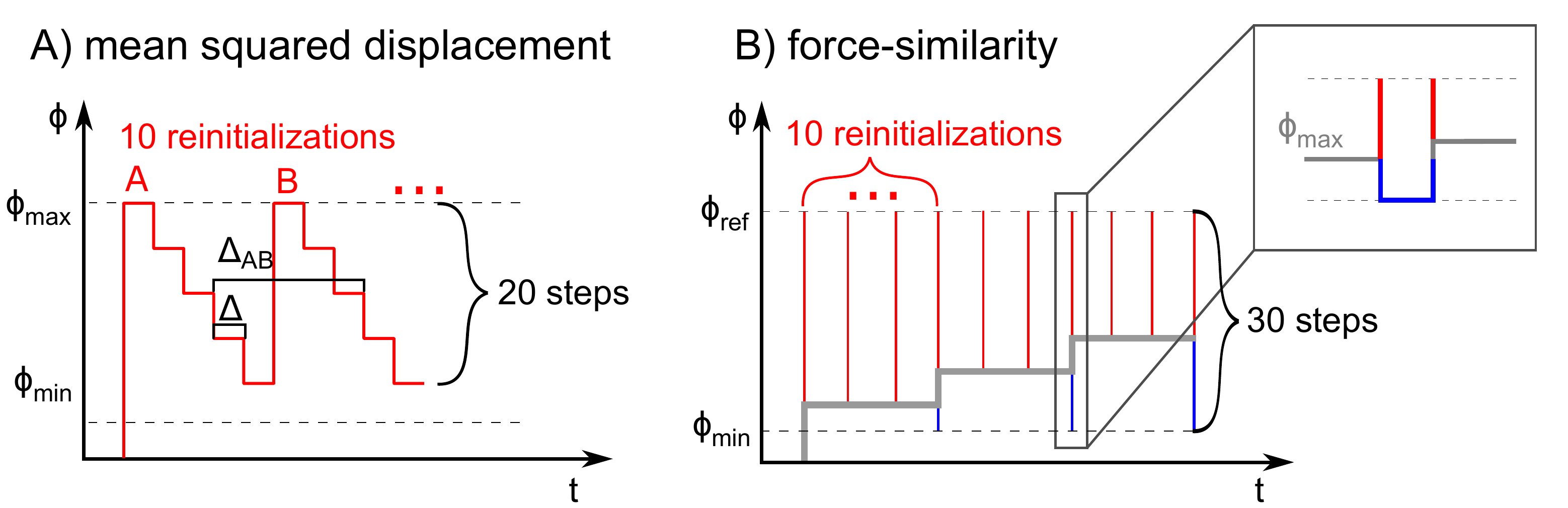}
    \caption{A) Schematic representation of the protocol used to determine the cage size $\Delta$, and the mean cage separation, $\Delta_{AB}$, via the mean squared displacement (MSD). B) Schematic representation of the protocol used to determine the persistence of inter-particle contacts, via force-similarity. 
    \label{fig:Methods}}
\end{figure*}

Figure~\ref{fig:Methods}A shows a schematic representation of the protocol used to locate the Gardner transition using the mean squared displacement (MSD). From $\phi_\mathrm{min}$, the system is first compressed to $\phi_\mathrm{max}$. The system is then decompressed stepwise, with $100t_{r} $ by airjet at each density. Upon reaching $\phi_\mathrm{min}$, the system is \textit{reinitialized} by compressing it to $\phi_\mathrm{max}$ again. The system is reinitialized a total of 10 times. The cage size $\Delta$ is calculated from the the ensemble of randomizations of a single initialization state of the system at a given $\phi$, and the cage separation distance $\Delta_{AB}$ by comparing particle positions of different initializations of the system at the same $\phi$.

Figure~\ref{fig:Methods}B shows a schematic representation of the protocol used to determine the evolution of the inter-particle forces as function of $\phi$. The system is first compressed to a jammed reference state $\phi_\mathrm{ref}$, to observe the photo-elastic fringes of the initial state, $\mathcal{I}$. The system is then decompressed to $\phi$ and let to evolve for $10 t_r$. The system is then recompressed to $\phi_\mathrm{ref}$, to observe the photo-elastic fringes of the final state, $\mathcal{F}$. These steps are repeated for 30 equispaced $\phi$, with $\phi_\mathrm{min} \leq \phi \leq \phi_\mathrm{max}$. Before advancing to the next $\phi$, the system is briefly brought to $\phi_{min}$ (highlighted by the inset), in order to limit the effect of aging. 

Note that, throughout the experiments, all values of $\phi$ are determined through \emph{a posteriori} image analysis, and thus do not necessarily coincide from one experimental setup to the other. The minimal values for the MSD measurements, $\phi_\mathrm{min} = 0.8006$ and for force similarity measurements, $\phi_\mathrm{min} = 0.8002$, are nevertheless identical within measurement error. The maximal measured values for the MSD measurements, $\phi_\mathrm{max} = 0.8162$ is, however, larger than for the force similarity measurements $\phi_\mathrm{max}=0.8127$. This choice makes the changes in the fringes more apparent, as forces become more homogeneous at even higher values of $\phi$. Because both $\phi_\mathrm{max}$ determined a jamming density  $\phi_\mathrm{J} = 0.8100$, at which rearrangements are not happening, the difference is not deemed significant.

\newpage
\subsection{Orthogonalizing lattice vectors}

\begin{figure*}
    \includegraphics[width=\textwidth]{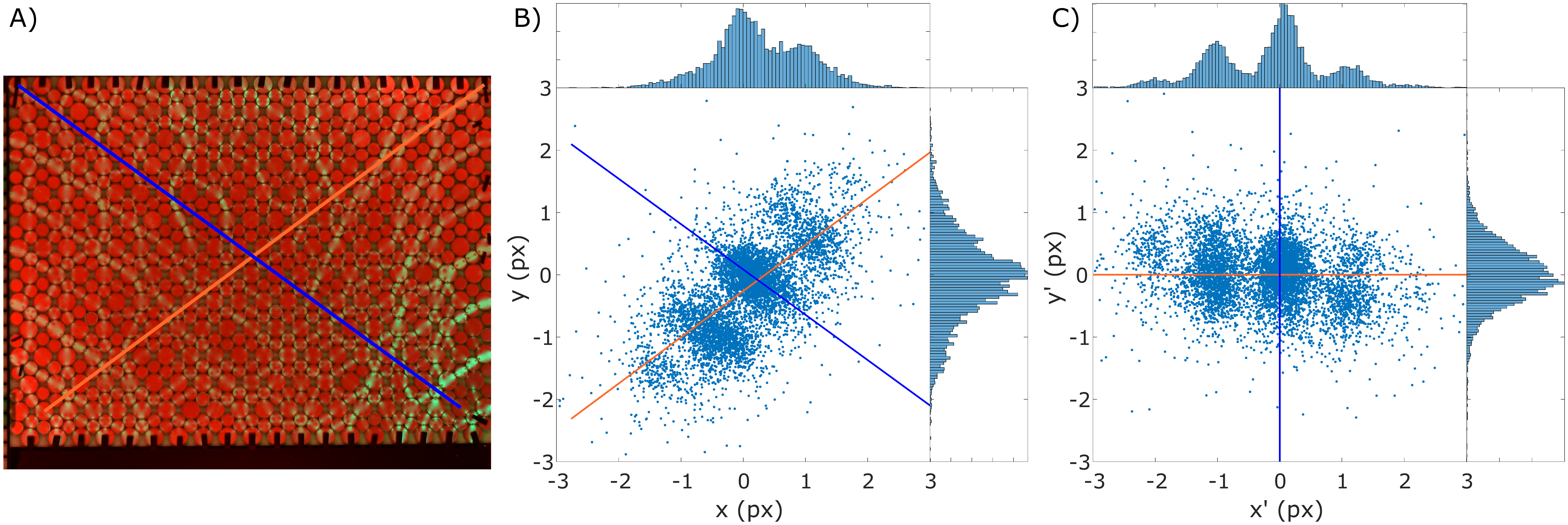}
    \caption{A) Typical image obtained during experiments, where the blue and orange lines indicate the lattice vectors of the crystal. B) Displacement components of $\Delta$ in the Gardner phase, with the blue and orange line indicating the lattice vectors of the crystal. The probability density functions in the $x$ and $y$ directions are given along their respective axis. C) Displacement components of $\Delta$ in the Gardner phase after linear transformation to orthogonalize the system, with the blue and orange line showing the lattice vectors of the orthogonalized system. The probability density functions in the $x'$ and  $y'$ directions of the orthogonalized system are given along their respective axis.}
    \label{fig:Delta_correction}
\end{figure*}

Recall that the H1 unit cell has lattice vectors which are both non-orthogonal to each other, and not perpendicular to the axes of the experiment, as shown in Fig.~\ref{fig:Delta_correction}A.
We observed that within the Gardner phase ($\phi \gtrsim 0.807$), the displacement components were no longer Gaussian and azimuthally symmetric around $v_x = v_y = 0$ (the mean however, was still at $v_x = v_y = 0$), as shown in Fig.~\ref{fig:Delta_correction}B. Instead, the displacement components split up into distinct bands, aligned with the lattice vectors of the crystal. 

Therefore, we applied a linear transformation to the data such that the lattice vectors became orthogonal; this resulted in two different probability density functions for the displacement components, shown in Fig.~\ref{fig:Delta_correction}C. The PDF in the $y'$-direction (blue axis) is now Gaussian, whereas the  distribution in the $x'$-direction is more complicated, and metrics like mean displacement, mean squared displacement, or standard deviation do not reflect the nature of the distribution, nor were they very reproducible. These features likely reflected an anisotropy in our experimental setup, that affected the configurational sampling of our system.  We therefore chose to use the distribution in the $x'$-axis to calculate $\Delta$, which showed no such artefact.

\end{document}